\documentstyle[twocolumn,prb,aps]{revtex}
\begin{document}
\input{psfig}
\twocolumn[\hsize\textwidth\columnwidth\hsize\csname @twocolumnfalse\endcsname
\title{Variational approach to protein design and extraction of
interaction potentials}
\author{Flavio Seno$^1$, Cristian Micheletti$^2$, Amos Maritan$^2$ 
and Jayanth R. Banavar$^3$}
\vskip 0.3cm
\address{(1) Dipartimento di Fisica, Universit\`a di Padova, 
Via Marzolo 8,35131  Padova, Italy and INFM}
\address{(2) International School for Advanced Studies (S.I.S.S.A.),
Via Beirut 2-4, 34014 Trieste, Italy and INFM}
\address{(3) Department of Physics and Center for Materials Physics,
104 Davey Laboratory, The Pennsylvania State University, University
Park, Pennsylvania 16802}
\date{\today}
\maketitle
\begin{abstract}
We present and discuss a novel approach to the direct and inverse
protein folding problem. The proposed strategy is based on a
variational approach that allows the simultaneous extraction of amino
acid interactions and the low-temperature free energy of sequences of
amino acids.  The knowledge-based technique is simple and
straightforward to implement even for realistic off-lattice proteins
because it does not entail threading-like procedures.  Its validity is
assessed in the context of a lattice model by means of a variety of
stringent checks.
\end{abstract}
\pacs{87.15.by,87.10.+e,64.60.Cn}
]
Two long standing challenges in molecular biology are the direct and
inverse problems of protein folding\cite{1}. The first deals with
the determination of the thermodynamically stable conformation of a
known sequence of amino acids \cite{MC,OLW} while the second involves
the elucidation of the amino acid sequence (if any) which admits a
given target structure as its stable conformation
\cite{DM97,D96,IP,hsf,M98a,M98b}.  One route to a solution of the direct
problem requires knowledge of the interaction potentials between the
protein constituents -- in principle one studies the energies of the
sequence in various conformations and identifies the native state
structure as being the one with the lowest energy \cite{MC}.  A
solution of the inverse or the sequence design problem entails
knowledge 
of the free energies of the  sequences
\cite{D96}.  This follows from an application of Boltzmann statistics
-- what matters is not how low the energy of a  sequence is
in the target conformation (a measure of which can be obtained from
the knowledge of the interaction potentials) but whether this energy
is lower than those in alternative conformations.
 
In this report, we introduce a variational approach for extracting the
interaction potentials between the protein constituents and the free
energies of candidate sequences simultaneously.  The method is general
and  applicable to real proteins. The input is a set of
sequences and their respective native structures, as available from
the Protein Data Bank. A  feature of the technique is that
alternative conformations that compete  with the native
state in housing a given sequence are not required. Here we apply this
method to a  lattice model of proteins to provide a stringent
validation of the approach.  Unlike real proteins, the interaction
potentials in such a model system can be chosen  and one may measure 
the accuracy with which these potentials
are determined as well as the effectiveness of sequence design.

We adopt the  approach of treating proteins at a
mesoscopic level with the amino acids being the fundamental units.
The influence of the internal degrees of freedom associated with each
amino acid as well as the solvent degrees of freedom are incorporated
through a coarse-grained Hamiltonian with effective interactions
between the amino acids.

The free energy, $F(S)$, of a sequence, $S$, is defined from the equation
\begin{equation}
e^{- \beta F(S)} =\sum_\Gamma e^{-\beta{\cal H}(S,\Gamma)} \ ,
\label{eqn:free}
\end{equation}
\noindent where ${\cal H}(S,\Gamma)$ is the energy of $S$ mounted on a
conformation $\Gamma$ and the sum is taken over all conformations that
the sequence can adopt.  A rigorous solution \cite{D96} of
the design problem on a target structure $\Gamma$ entails the
identification of the sequence(s), $S$, that maximizes the functional

\begin{equation}
P_\Gamma(S) = e^{- \beta ({\cal H}(S,\Gamma) - F(S)) } \ ,
\label{eqn:func}
\end{equation}

\noindent evaluated at low temperature ( below the
folding transition temperature where $P_\Gamma(S)=1/2$).
 $P_\Gamma(S)$ is the probability that a sequence $S$ is found in
conformation $\Gamma$ at an inverse temperature $\beta$.  Thus the
solution $S$  is the sequence which has the
highest low-temperature probability of being found on $\Gamma$.  At
low temperatures \cite{temp}, a sequence $\bar{S}$ with a {\em unique}
ground state, $\bar{\Gamma}$, satisfies the inequality

\begin{equation}
H(\bar{S},\bar{\Gamma}) - F(\bar{S}) \le H({S},\bar{\Gamma}) - F({S}),
\label{eqn:ineq}
\end{equation}

\noindent for arbitrary sequence, $S$, with the equality possibly
holding only when $S$ admits $\bar{\Gamma}$ as its native state. A range
of such equalities could be used to determine optimal values of
variational parameters characterizing the interactions and the low temperature
free energies.

The maximization of $P_\Gamma(S)$ is computationally demanding because
it involves the calculation of $F(S)$ for each  amino acid
sequence and  an exact calculation of $F(S)$ for a given sequence $S$
involves a sum over the enormous number of its possible conformations.
The use of importance
sampling techniques for the estimation of $F(S)$ at low-$T$
requires efficient algorithms to find 
conformations that compete significantly with $\Gamma$\cite{D96}.  
Such an approach has been used
fruitfully for lattice models of proteins \cite{M98a}, but is not
 feasible for realistic off-lattice cases \cite{M98b}.

$F(S)$ formally depends only on $S$ and hence one may postulate a
functional form of $F$ which depends  on sequence properties (e.g.
 the concentration of amino acids) \cite{M98a,M98b}.  At $T=0$
\cite{temp},
 the free energy of a sequence 
ought to be exactly equal to its energy in
the native state conformation (which  depends on the
conformation and the interaction potentials) -- this forms the basis
for our variational approach. Unlike the inequalities (\ref{eqn:ineq}),
the new approach does not entail the mounting of a sequence on any but
its own native state conformation. We define an intensive functional
$\Delta$ (whose choice is not unique), whose minimization can be used
to identify a consistent set of potential and free energy parameters.
A convenient choice that we  used in our  calculations is

\begin{eqnarray}
&&\Delta =  \bar{\epsilon} \biggl\{ \sum_i\biggl( { {\cal
H}(S_i,\Gamma_i) - F(S_i)  \over 
L_i} \biggr)^2 \nonumber \\ 
&+&\biggl( { {\cal H}(S_i,\Gamma_i) - F(S_i)  \over
L_i} \biggr)^4 \cdot \Theta[ F(S_i) - {\cal H}(S_i,\Gamma_i)]  
\biggr\} \ ,
\label{eqn:delta2} 
\end{eqnarray}

\noindent where $\Theta[x]$ is the Heavside  function,
and the sum is taken over the sequence-native state
conformation set in the protein data bank, and $L_i$ is the length of
the $i$th sequence.  The second term in (\ref{eqn:delta2}) is used to
penalize cases for which the parameters violate the physical
constraint, ${\cal H}(S_i,\Gamma_i) \ge F(S_i)$.
The quantity $\bar{\epsilon}$ is the absolute value of the
average of the interaction strengths between amino acids and its
utility is explained below.  A zero value for $\Delta$ would
correspond to a perfect parametrization of both the interaction
potentials and the free energies for the finite set of sequences in
the data bank.  More generally, for a finite protein data bank, there
will exist a non-zero region in the parameter space of potentials and
free energies within which $\Delta$ is at a minimum. With perfect
parametrization, this region would be expected to shrink around the
 parameter values as the data bank size increases
\cite{VM98}. It should be stressed that, contrary to common potential
extraction or design procedures, the minimization of the functional
(\ref{eqn:delta2}) does {\em not} involve the use of decoy structures
nor the mounting of sequence $i$ on any structure other than its
ground state, $\Gamma_i$. In order to create a data bank, a random
exploration of the ensemble of 4-amino-acid sequences of length 16 was
performed to select those admitting a unique ground state
conformation. The possible protein conformations were assumed to be
self-avoiding oriented walks embedded on a square lattice \cite{3}
with an interaction between amino acids $i$ and $j$ only if they are
next to each other on the lattice and yet not next to each other along the
sequence. 

We chose an interaction matrix, $\epsilon$, between the 4
different types (or classes) of amino acids. These are the entries of
the 4x4 $\epsilon$-matrix in the first row of Table \ref{table1} (with
$\epsilon_{1,1}=-40$).

 To mimick the thermodynamic
stability of proteins, we further
selected the sequences and retained only those with an energy gap
between the unique native state and first excited state energies $\ge 10$, a
constraint satisfied by, roughly, 1\% of the sequences.  Our final 
data bank consisted of 500 sequences with
their  ground state, $\{ S_i, \Gamma_i \}_{i=1 ... 500}$.

In our model studies, we chose to parametrize the interaction matrix
with the same functional form as the true interaction matrix but with
nine variational parameters in the symmetric $\epsilon$ matrix 
($\epsilon_{1,1}$ was held fixed at a value
of -40 in order to set the energy scale).  We assumed the simplest
form for an extensive free energy \cite{M98a,M98b} with four variational
parameters (denoted by $a_i$, $i=1...4$):

\begin{equation}
F(S) = a_1 \cdot n_1 + a_2 \cdot n_2 +a_3 \cdot n_3 +a_4 \cdot n_4 \ ,
\label{eqn:free0}
\end{equation}

\noindent where $n_i$ is the number of amino acids of type $i$ found
in $S$.  Eq. (\ref{eqn:free0}) may be viewed as the lowest order 
expansion of $F$ in the ``order parameters'', $n_i$'s.

$\Delta$ was minimized using a simulated annealing procedure with the
constraint 
$\epsilon_{1,1} \leq \epsilon_{i,j} < 0 $.
The hierarchy of amino acids
interaction strengths was mirrored by the frequencies of pair contacts
in the data bank and allowed for a restriction of the search in
parameter space.

The quantity $\bar{\epsilon}$ in (\ref{eqn:delta2}) was useful in
avoiding convergence to a spurious trivial solution in which all the
$\epsilon_{i,j}$'s are equal to $\epsilon_{1,1}=-40$, and $F(S)$
becomes (-40) times the number of contacts.

The functional $\Delta$ was minimized using subsets of our global data
bank within which the number of elements ranged from 100 to 500. The
extracted potentials, as well as the free energy coefficients appear
in Table \ref{table1}. We further checked, using the extracted
parameters, whether each sequence in the data bank recognized the
associated structure as its ground state among all the possible
conformations.  The success rate was typically $>$ 80\% with an
increase in the success rate on increasing the size of the data bank.

We then proceeded to use the functional $(H-F)$ to carry out sequence
design on a target structure.  This entails the identification from
among the $4^{16}$ sequences the one that minimizes $(H-F)$ (using the
extracted parameters) on the target structure $\Gamma$.  The
correctness of the design is checked by using the true Hamiltonian to
verify whether the designed sequence admits $\Gamma$ as its (possibly
degenerate) ground state.  Our test was performed on 100 structures
taken from our data bank using a Monte Carlo procedure.

\begin{figure}
\vskip 0.5cm
\centerline{\psfig{figure=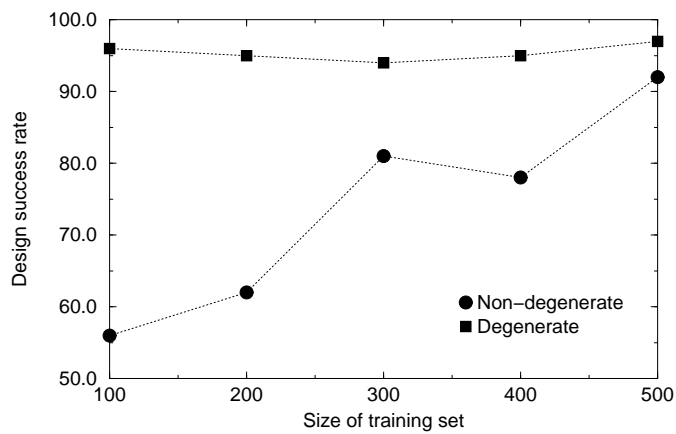,width=3.5in}}
\vskip 0.5cm
\caption{Plot of the success rate in identifying the sequence that
admits a pre-assigned target structure as its degenerate (squares) and
non-degenerate (circles) ground state as a function of the training
set size. The results were obtained with a single randomly chosen set
of each size.}
\label{fig:design}
\end{figure}

Fig. \ref{fig:design} shows a plot of the design success rate as a
function of the size of the training set.  It is worth noting that
none of the designed sequences appeared in the original data bank. Our
analysis was not limited to those sequences with the lowest $(H-F)$
score; we extended it to the 10 highest ranking sequences for each
target structure. Using the parameters deduced from the training set
of size 500, we found an excellent overall design success of 88\% and
92\% for unique and degenerate encoding respectively.

\begin{figure}
\vskip 0.5cm
\centerline{\psfig{figure=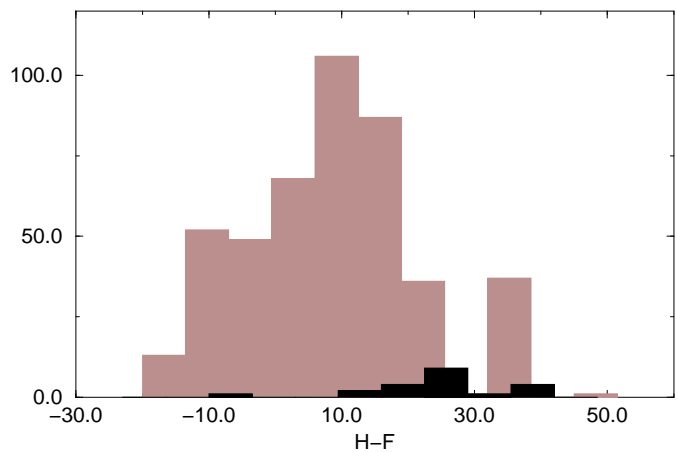,width=3.5in}}
\vskip 0.5cm
\caption{Distribution of the quantity $(H-F)$ for the correctly chosen
sequences (gray) and the improper sequences (black).}
\label{fig:histogram}
\end{figure}

\begin{figure}
\vskip 0.5cm
\centerline{\psfig{figure=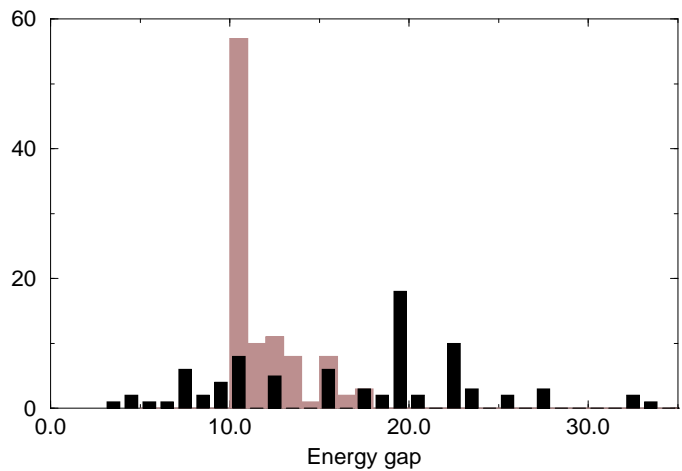,width=3.5in}}
\vskip 0.5cm
\caption{Distribution of the energy gaps between the native state
and first excited state energies for the sequences in the data
bank (gray) and designed sequences that have a non-degenerate
ground state (black).}
\label{fig:gaps}
\end{figure}

In Fig. \ref{fig:histogram} we have plotted the histogram of the
$(H-F)$ distribution for the improperly chosen sequences (black) and
the correct ones (gray). The $(H-F)$ score for the improperly chosen
sequences take on large positive values, signalling that the estimated
energy, $F$ of the sequence in its unknown native state is
substantially lower than in the target structure. Thus one may discard
{\em a priori} the majority of bad solutions by a mere inspection of
their large $(H-F)$ scores.  The unphysical negative values of $(H-F)$
originate from the small size of the training set and the imperfect
parametrization of the free energy. We also considered several
generalizations of (\ref{eqn:free0}) including 
two-body terms of the form $n_i \cdot n_j$ and chemical potentials
that control the number of ``walls'' separating segments of identical
amino acids \cite{M98b} with slight improvement in the success rates.
A further check of the quality of the designed sequences was performed
by inspecting the distribution of their energy gaps  versus 
those used in the data bank.  The designed sequences
tend to have energy gaps between the native state and the first
excited state that are larger than those of sequences in the data bank
(Figure \ref{fig:gaps}) showing that the design procedure yields
sequences with a higher thermodynamic stability.

Finally, we performed a challenging blind test to assess the validity
of the variational approach.  The coefficients extracted for the
16-bead case were used to carry out sequence design on a compact
target conformation of length 25.  The target conformation was
RRDLDRDDLULDLLURULUURDRD (where R,L,U,D stand for right, left, up and
down respectively and indicate the directions of the bonds that define
the self-avoiding lattice conformation) which is highly designable
\cite{Li2} in a different lattice model \cite{M98c} and is
geometrically regular \cite{Li2}. The sequence chosen was
124211324211324211324211. Indeed, an exhaustive search of the native
state of the sequence among all self-avoiding walks of length 25
confirmed that this sequence had the target structure as its unique
ground state.

In order to ensure that the strategy used here is robust and
independent of the particular choice of the $\epsilon$ matrix and/or
data bank, we performed a similar analysis using another randomly
generated interaction matrix and found results of statistically
similar quality as summarized in the bottom of Table \ref{table1}.

Our results show that one may define a design score, $\Delta$, that
takes on small values for sequences mounted on their true native state
and large positive values for improper mounting. It is striking that
the simple free energy form as in (\ref{eqn:free0}) can be so
effective for building a reliable $\Delta$ functional. A physically
appealing explanation for this is to regard the parameters in
(\ref{eqn:free0}) as controlling both the residue composition of the
designed sequence as well as indicating its expected ground state
energy. The solution to a design problem will be provided by the
sequence(s) that meets the composition requirements and which, when
mounted on the target structure, has an energy equal to or better than
the expected value. Thus, the variational approach provides a feedback
mechanism for design; it is self-regulating in that no external action
is required to rule out runaway solutions favouring the abundance of
the most energetically favoured contacts.  This self-regulating
mechanism also counterbalances an improper parametrization of $H$
and/or $F$, thus decreasing the sensitivity of the overall ($H-F$)
score to the detailed functional form of $\Delta$.

In conclusion, we have presented a novel procedure for tackling the
direct and inverse folding problems simultaneously. The proposed
strategy is general and ought to be applicable to the case of real
proteins. We have discussed a practical implementation of the
technique and have carried out rigorous testing of its efficiency in
folding and design.  The results are encouraging and are suggestive of
the feasibility of a simple parametrization of the free energy of
sequences of amino acids.

This work was supported in part by INFN sez. di Trieste, NASA, NATO
and the Center for Academic Computing at Penn State.


\twocolumn[\hsize\textwidth\columnwidth\hsize\csname @twocolumnfalse\endcsname
\begin{table}
\begin{center}
  \vskip 0.5cm
\begin{tabular}{|r|r|r|r|r|r|r|r|r|r|r|r|r|r|} \hline
PDB Size & $\epsilon_{1,2}$ & $\epsilon_{1,3}$ & $\epsilon_{1,4}$
& $\epsilon_{2,2}$ & $\epsilon_{2,3}$ & $\epsilon_{2,4}$ &
$\epsilon_{3,3}$ & $\epsilon_{3,4}$ & $\epsilon_{4,4}$ &$a_1$&
$a_2$&$a_3$&$a_4$ \\ \hline\hline
TRUE & -30 & -20 & -17 & -25 &-13& -10  & -5 & -2 & -1 & & & & \\ \hline
100 &   -32.63&  -26.80&  -22.71&  -31.57&  -22.71&  -17.76&  -17.76&
-17.75&   -0.00&  -26.50&  -17.44&  -12.60&  -10.39 \\
200&  -32.62&  -25.56&  -23.17&  -30.65&  -23.17&  -16.06&  -14.07&
-5.43&   -0.00&  -26.15&  -17.14&  -12.33&  -11.07\\ 
300        &  -31.27&  -27.91&  -23.29&  -27.91&  -23.17&
-12.61&   -8.50&   -8.50&   -0.00&  -27.17&  -15.62&  -11.83&
-10.87\\ 
400        &  -31.03&  -27.14&  -23.27&  -27.14&  -22.44&
-12.50&   -6.94&   -6.77&   -0.00&  -27.62&  -15.25&  -11.52&
-10.38\\ 
500&  -32.23&  -25.93&  -24.12&  -28.55&  -22.78&  -16.40&  -11.79&
-9.13&   -5.21&  -25.63&  -16.49&  -10.53&   -9.55\\ \hline \hline 
TRUE & -22 & -18&  -12 &-11&  -17 &  -1& -28 & -13 & -1 & & & & \\ \hline
250&  -24.05&  -17.95&  -13.22&  -13.02&  -17.95&   -0.00&  -24.06&
-13.22&   -0.00&  -26.26&  -10.41&  -12.70&   -4.14  
\end{tabular}
\vskip 0.5cm
\caption{A summary of the results with two data banks containing 500
and 250 training proteins respectively. 
In all cases $\epsilon_{1,1}$ was fixed at
-40 in order to set the energy scale. The row entitled TRUE shows the
true potential parameters in both cases.  The other rows show the
values of the extracted parameters of the potential and the free
energy with the number in the first column showing the number of
proteins in the training set. A single randomly chosen set was
employed in each case. For the second data bank, the folding success
rate was 91 \%, while the unique and degenerate design success rates
were 73\% and 96\% respectively.}
\label{table1}
\end{center}
\end{table}  
]


\begin{references}  
\bibitem{1} C. Branden, \& J. Tooze, (1991) in {\em Introduction to
protein structure}, Garland Publishing, New York; T.E. Creighton,
(1992) in {\em Proteins: structures and molecular properties}.



\bibitem{MC} V. N. Maiorov and G. M. Crippen, {\em J. Mol. Biol.},
{\bf 227}, 876 (1992); S. Miyazawa and R. L. Jernigan, {\em Macromolecules}, {\bf 18}, 534 (1985); {\em J. Mol. Biol.} {\bf 256},
623 (1996); M. J. Sippl and S. Weitckus, {\em Proteins: Structure
Function and Genetics}, {\bf 13}, 158 (1992).


\bibitem{OLW} J. N. Onuchic,Z. LutheySchulten and P. G. Wolynes, {\em
Ann. Rev. of Phys. Chem.} {\bf 48}, 545 (1997); J. N. Onuchic,
N. D. Socci, Z. LutheySchulten and P.G. Wolynes {\em Folding and
Design} {\bf 1}, 441 (1996); E. I.  Shakhnovich {\em Folding and
Design} {\bf 1}, R50 (1996); L. A. Mirny and E. I. Shakhnovich {\em
J. Mol. Biol.} {\bf 264}, 1164 (1996); T. Veitshans, D. Klimov and
D. Thirumalai, {\em Folding and Design} {\bf 2}, 1 (1997); G. D. Rose
and R. Srinivasan {\em Biophys. Journ.} {\bf 70}, WPMS4 (1996).


\bibitem{DM97} B. I. Dahiyat and S. L. Mayo, {\em Science} {\bf 278},
82 (1997).


\bibitem{D96} J. M. Deutsch and T. Kurosky, {\em Phys. Rev. Lett} {\bf
  76}, 323-326 (1996); F. Seno, M. Vendruscolo, A. Maritan,
  J. R. Banavar, {\em Phys. Rev. Lett.} {\bf 77}, 1901-1904 (1996);
  M. P. Morrissey and E. I. Shakhnovich, {\em Folding and Design} {\bf
  1}, 391 (1996).

\bibitem{IP}  A. Irb\"ack, C. Peterson, F. Potthast, E. Sandelin,
preprint, cond-mat/9711092. 


\bibitem{hsf} Yue, K.,  Fiebig, K. M.,  Thomas, P. D.,  Chan, H. S., 
Shackhnovich, E. I.,  Dill, K. A., 
{\em Proc. Natl. Acad. Sci. USA} {\bf 92}, 325-329 (1995).


\bibitem{M98a} C. Micheletti, F. Seno, A. Maritan and J. R. Banavar,
{\em Phys. Rev. Lett.} {\bf 80}, 2237 (1998)

\bibitem{M98b} C. Micheletti, F. Seno, A. Maritan and J. R. Banavar,
{\em Proteins: Struct. Funct. and Gen}, in press,
cond-mat/9712124.

\bibitem{temp} The low-temperature regime considered here includes
room temperature in protein experiments and so thermal fluctuations 
 do not alter the protein structure in its native
state.



\bibitem{VM98}  J. van Mourik, C. Clementi, A. Maritan, F. Seno,
J. R. Banavar,
preprint, cond-mat/9801137

\bibitem{3} K. F. Lau and K. A. Dill, {\em Macromolecules} {\bf 22},
3986-3997 (1989); H.S.  Chan and K. A. Dill, {\em Physics Today}, {\bf
46}, 24 (1993); K. A. Dill, S.  Bromberg, S. Yue, K.  Fiebig, K. M.
Yee, P. D. Thomas and H. S. Chan, {\em Protein Science} {\bf 4}, 561
(1995); P.D. Thomas and K.A. Dill, {\em Proc. Natl. Acad.  Sci. USA},
{\bf 93}, 11628 (1996).

\bibitem{Li2} H. Li, R. Helling, C. Tang, N. Wingreen, {\em Science}
{\bf 273}, 666 (1996).


\bibitem{M98c} C. Micheletti, J. R. Banavar,  A. Maritan and  F. Seno,
preprint, cond-mat/9712277. 

 
\end{references}
\end{document}